\documentclass[aps,pra,preprint,showpacs]{revtex4-1}
\usepackage{graphicx,color}
\usepackage{amsmath}
\usepackage{braket}
\usepackage[percent]{overpic}

\def\red#1{\textcolor{red}{#1}}

\begin{document}

\title{Vortex scattering by impurities in a Bose-Einstein condensate}
\author{A. Griffin$^{1,2}$}

\author{G. W. Stagg$^1$}

\author{N. P. Proukakis$^1$}

\author{C. F. Barenghi$^1$}
\email{carlo.barenghi@newcastle.ac.uk}

\affiliation{$^1$Joint Quantum Centre (JQC) Durham--Newcastle, School of Mathematics and Statistics, Newcastle University, Newcastle upon Tyne, NE1 7RU, United Kingdom}

\affiliation{$^2$Mathematics Institute, University of Warwick, 
Coventry CV4 7AL, United Kingdom}

\date{\today}

\begin{abstract}
Understanding quantum dynamics in a two-dimensional Bose-Einstein
condensate (BEC) relies on understanding how vortices interact with 
each others microscopically and with local imperfections of the 
potential which confines the condensate. Within a system consisting of
many vortices, the trajectory of a 
vortex-antivortex pair is often scattered by a third vortex, 
an effect previously characterised. However, the natural question 
remains as to how much of this effect is due to the velocity 
induced by this third vortex and how much is due to the density inhomogeneity
which it introduces. 
In this work, we describe the various 
qualitative scenarios which occur when a vortex-antivortex pair interacts
with a smooth density impurity whose profile is identical to that of a 
vortex but lacks the circulation around it.

\end{abstract}

\pacs{}

\maketitle
    
  \section{Introduction}
In a recent paper \cite{Smirnov2015}, 
Smirnov \& Smirnov have studied the scattering of
two-dimensional (2D) 
vortex-antivortex pairs and solitons
by a single quantum vortex in a homogeneous
atomic Bose-Einstein condensate. They found that the pair is scattered
over large angles radiating sound waves, in agreement with earlier
calculations \cite{Parker2005}. This scattering process is important 
because it lies at the heart of the dynamics 
of 2D quantum turbulence, a problem which is
currently attracting experimental  and theoretical attention
\cite{Nazarenko2007,Neely2013,Kwon2014,Stagg2015,Cidrim2016,Groszek2016}. 
Our understanding of the turbulent motion of many interacting vortices is
based on recognizing the most elementary interactions, such as the
interaction of a vortex with another vortex of the same or opposite
sign (resulting respectively in rotational or translation motion of
the pair). Similarly, we would like to recognize the possible elementary
interactions between a vortex and a large density perturbation induced by
the dynamics of vortices by external means.
It is well-known that a quantum
vortex in a Bose-Einstein condensate is a hole of zero density 
around which the phase changes by $2 \pi$. 
The natural question is
whether the incoming vortex-antivortex pair would be scattered
(and if so, by which amount) by a density perturbation alone (without the
circulation around it), as density gradients induce a Magnus force
\cite{McGeeHolland2001,FetterSvidzinsky2001} which deflects the pair.
To answer this question, we have performed numerical simulations 
of vortex-antivortex pairs travelling towards a fixed target in the
form of a density perturbation (hereafter referred to as an `impurity') and
whose depth and size is similar to the depth and size of a quantum
vortex (but without the circulation). Here we report about the
significant scattering induced by the impurity, and compare it with
the scattering induced by a target in the form of a vortex.

For simplicity we consider
a homogeneous condensate at zero temperature, and aim at identifying
the various qualitative scenarios which are possible (quantitative
predictions of vortex trajectories in a harmonically trapped
condensate require more specific calculations which depend on the
actual physical parameters and geometry, and are outside the scope
of this work).  A better physical
understanding of the scattering which imperfections
induce on vortices is generally useful (although in general imperfections
may not be as symmetric as we describe them here).
Our results are also relevant to
the manipulation of vortices using optical potentials generated
by laser beams \cite{Davis2009,Aioi2011}.

\section{Model}

\begin{figure}
\centering
\includegraphics[width=0.3\linewidth]{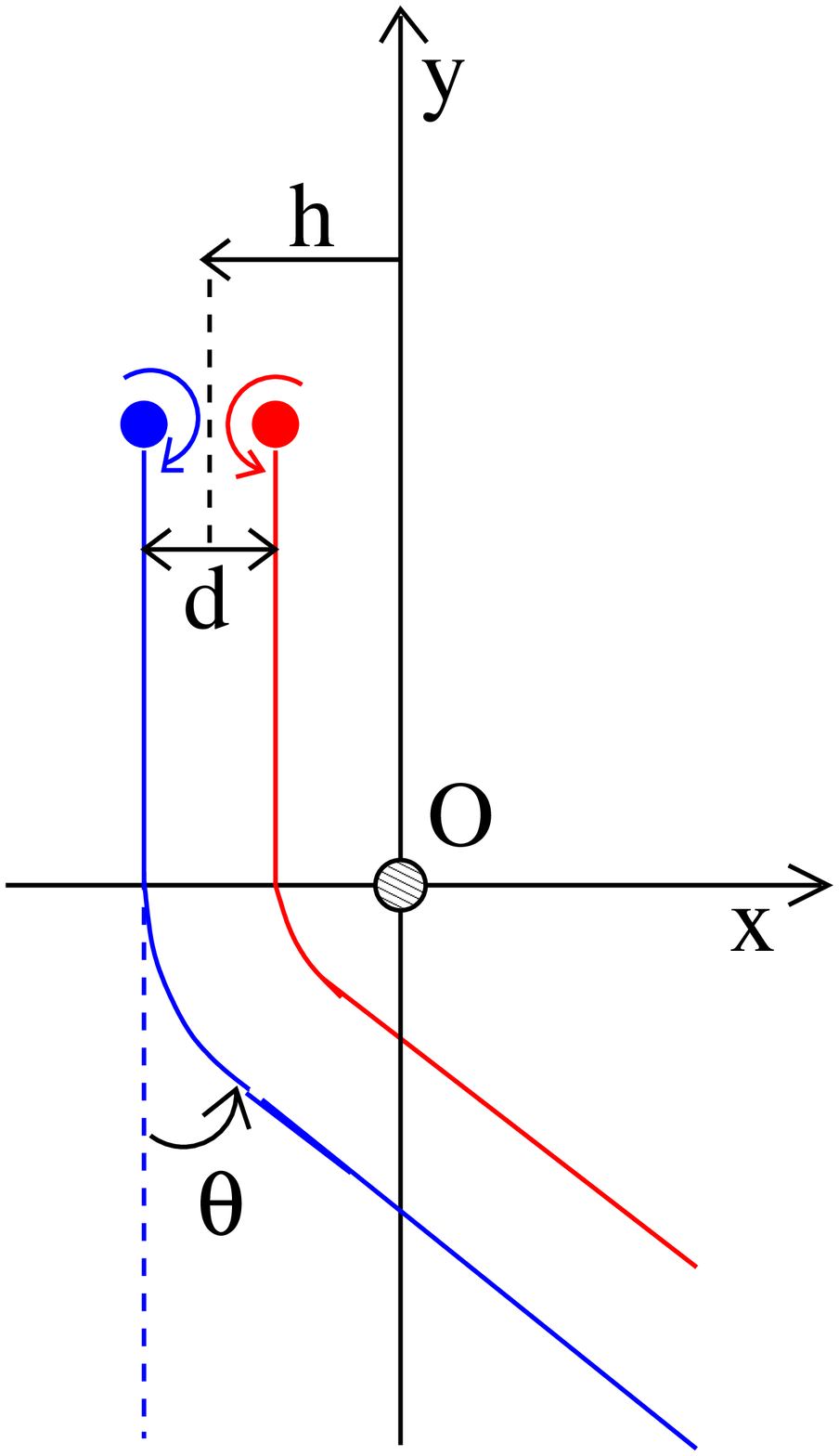}
\caption{(Color online). Schematic representation of the
scattering configuration. Initially, the vortex (red, right) and the
anti-vortex (blue, left) are separated by the distance $d$; the impact
parameter is $h$ (here $h<0$). The vortex-antivortex pair travels in the
negative y-direction towards the impurity (or a third vortex) at the
origin. The scattering angle is $\theta$ (here $\theta>0$).
}
\label{revfig1}
\end{figure}

Our model is the 2D Gross-Pitaevskii equation (GPE) 
for a homogeneous condensate at zero temperature. We use
dimensionless variables based on the healing length 
$\xi=\hbar/\sqrt{m \mu}$, the time scale $\tau=\hbar/\mu$ and the 
number density $n_0=\mu/g$, where $\mu$ is the chemical potential,
$g$ is the (2D) interaction parameter, $\hbar=h/(2 \pi)$ and $h$ is 
Planck's constant; the external potential $V$ is scaled by
$\mu$. The resulting dimensionless GPE 

\begin{equation}
i  \frac{\partial \psi }{\partial t}=
\left( -\frac{1}{2}\nabla^{2} + V +\left|\psi \right|^{2} - 1 \right ) \psi,
\label{eq:GPE}
\end{equation}
 
\noindent
is solved in the (dimensionless)
2D domain $-L \leq x,y \leq L$.
The initial condition at $t=0$, schematically described in Fig.~\ref{revfig1},
is a vortex-antivortex pair,
consisting of a left (clockwise) vortex and a right (anticlockwise) vortex
initially placed respectively at positions $x_L(0)$, $y_L(0)$ and
$x_R(0)$, $y_R(0)$.
We call $d=\vert x_L(0)-x_R(0)\vert$.
the initial distance between the vortices of the pair.
The vortex-antivortex pair travels along the negative $y$ direction 
with impact parameter $h = (x_R(0)+x_L(0))/2$
towards a fixed density perturbation (or impurity) held 
at $x_I=y_I=0$.
The impurity is represented by the (dimensionless) external potential
$V(x,y)=\sum_{j=1}^4 A_j e^{-R^2/\sigma_j^2}$ where
$R_j^2=((x-x_I)^2 + (y-y_I)^2)/\sigma_j^2$.
With a suitable choice of parameters $A_j$ \cite{parameters},
solving the time-independent GPE without the vortex-antivortex pair,
the density profile of the impurity approximately matches 
the profile of a singly-charged vortex at $x_I$, $y_I$ in the
homogeneous condensate.

To quantify the scattering,  we measure the deflection 
angle $\theta$ of the antivortex away
from its initial trajectory, but in some cases (for example if
the vortex-antivortex pair breaks up) a different description of 
the interaction is necessary.

We choose $L=76.65$ and impose $\psi=0$ on the boundaries.
We use a $1024^2$ grid,
corresponding to the (dimensionless)
spatial discretization $\Delta x=\Delta y=0.15$. 
Time-stepping is performed using the 4th-order Runge-Kutta scheme;
the typical (dimensionless) time step is $\Delta t= 0.01$.
During typical evolutions the total energy is
conserved within $0.003~ \%$. 
The calculations were repeated in a $512^2$ box with discretization
$\Delta x=\Delta y=0.3$, resulting in the same qualitative 
scattering scenarios which we describe in the next section. Variations in the
deflection angles resulting from discretization errors and from sound waves 
reflected from the boundaries are small (of the order of one percent);
the trapping scenario appears more sensitive to perturbations.
On the other hand, in the experiments, 2D vortex configurations typically 
contain significant sound waves besides thermal noise.


\section{Scattering scenarios}
\label{sec:results}

\begin{figure}[ht]
\centering
  \includegraphics[width=0.18\linewidth]{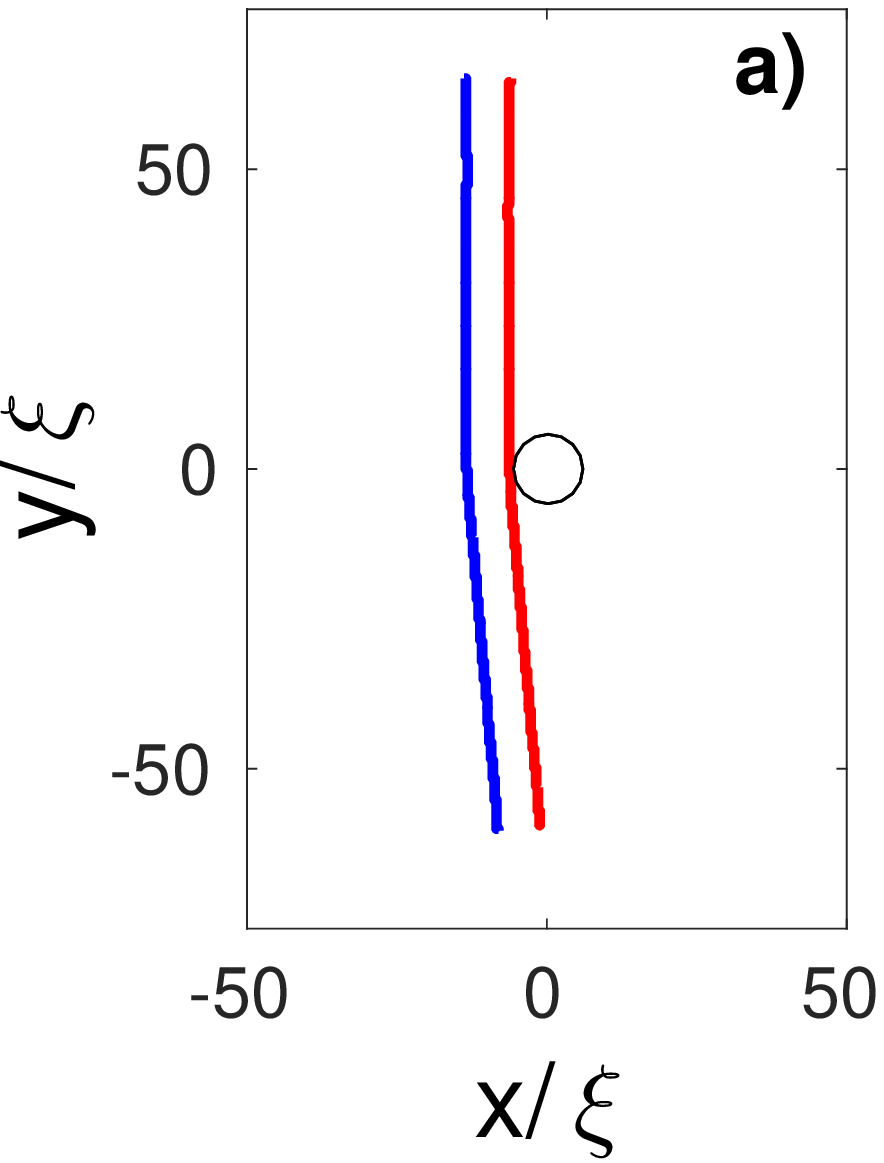}
  \includegraphics[width=0.18\linewidth]{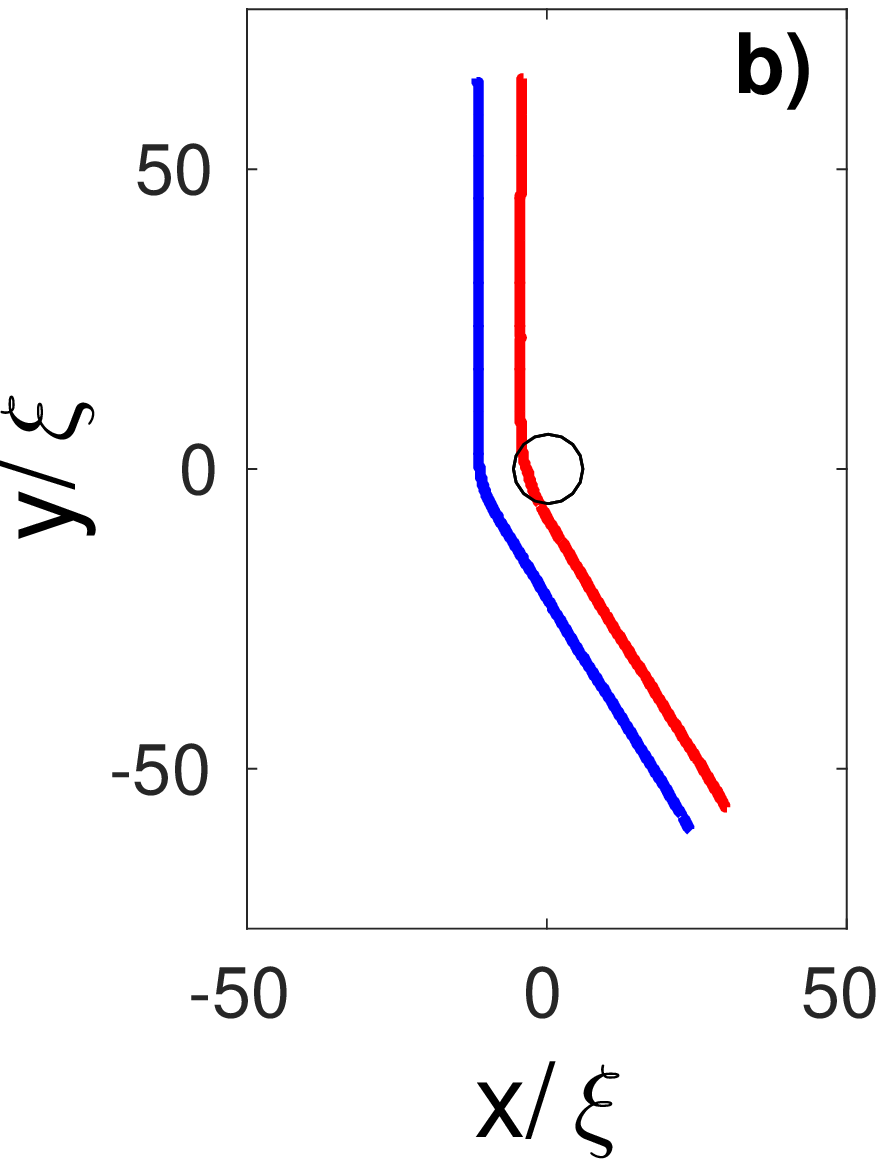}
  \includegraphics[width=0.18\linewidth]{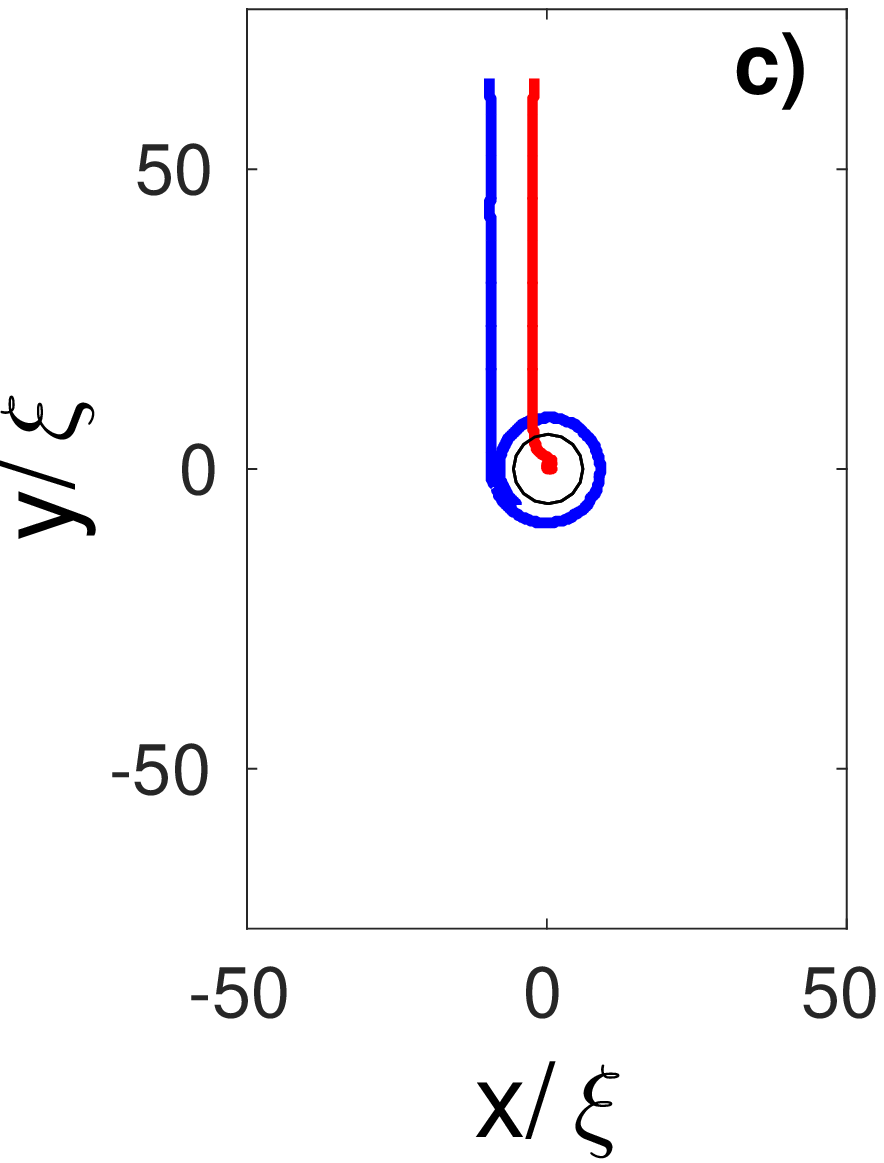}
  \includegraphics[width=0.18\linewidth]{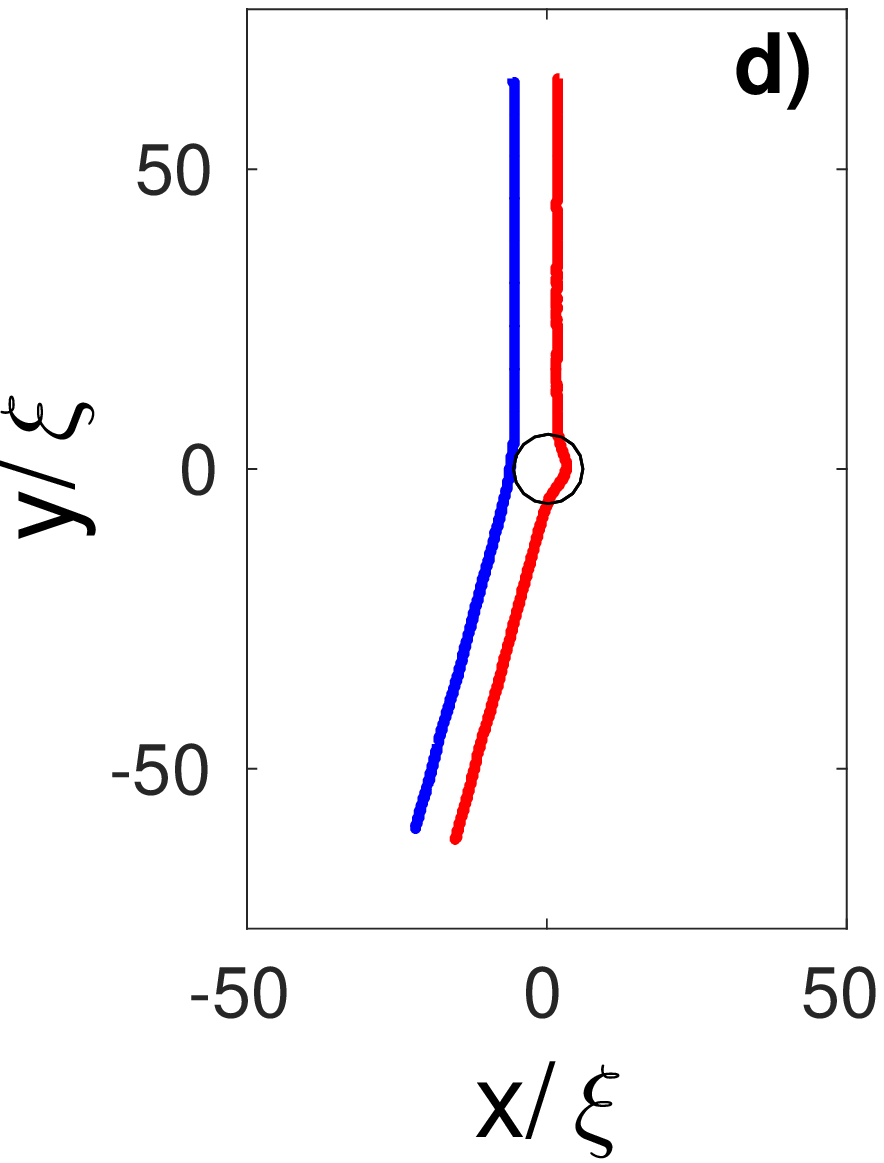}
  \includegraphics[width=0.18\linewidth]{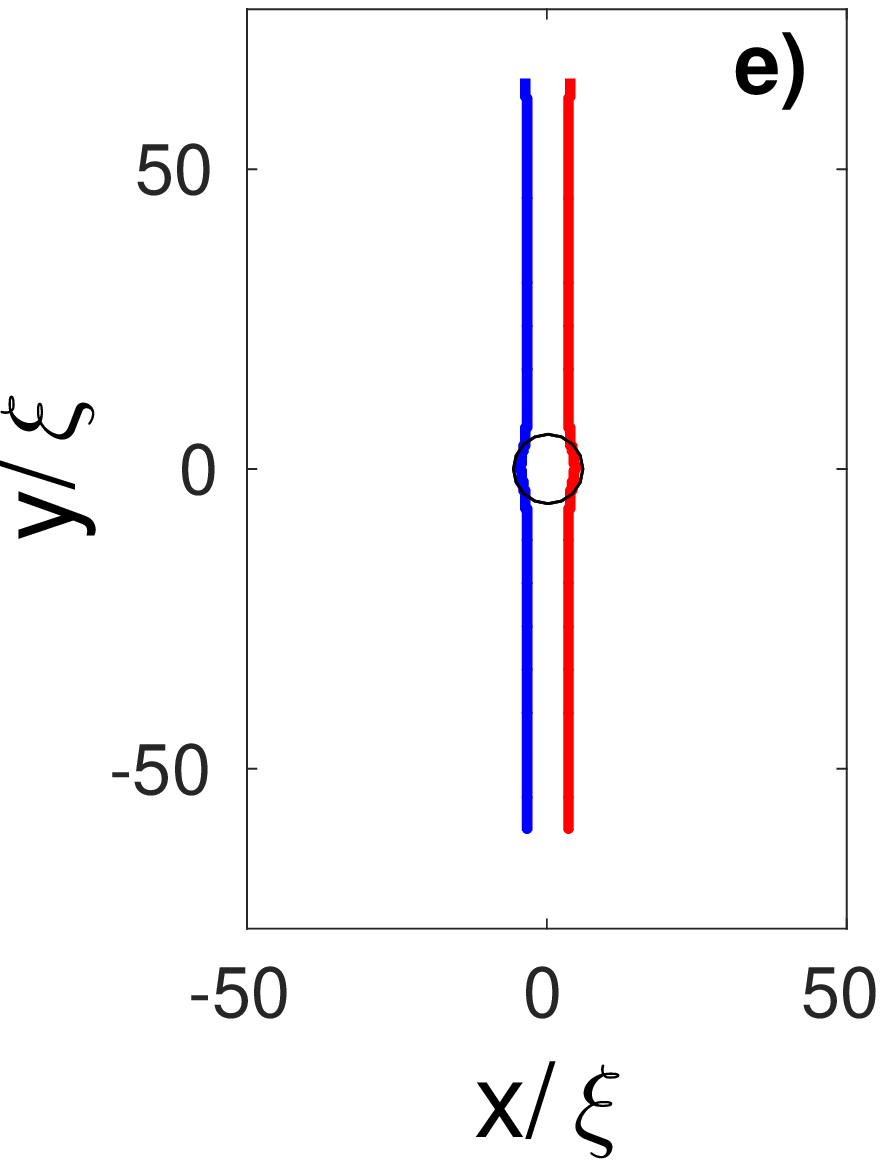}\\
  \includegraphics[width=0.18\linewidth]{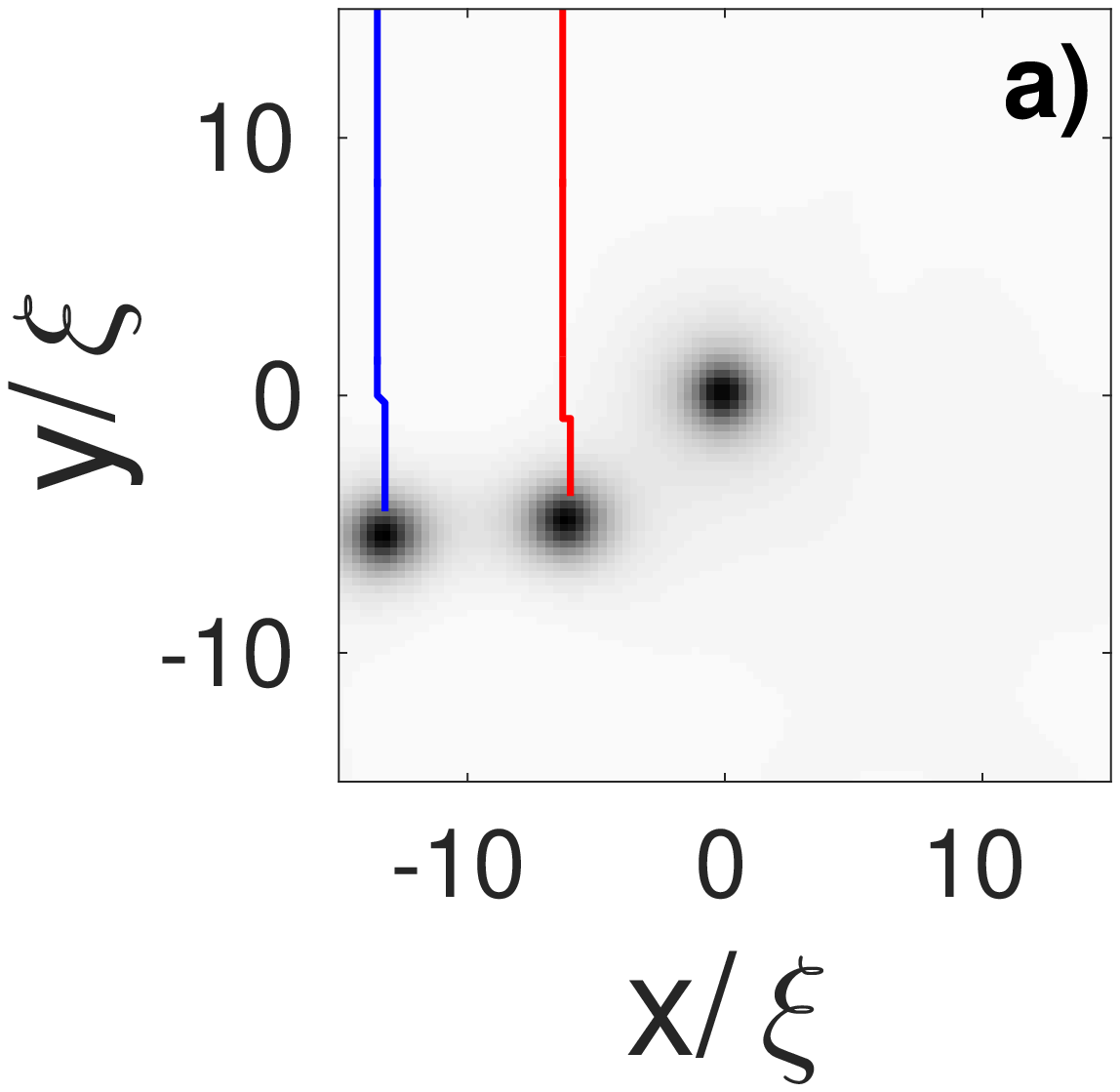}
  \includegraphics[width=0.18\linewidth]{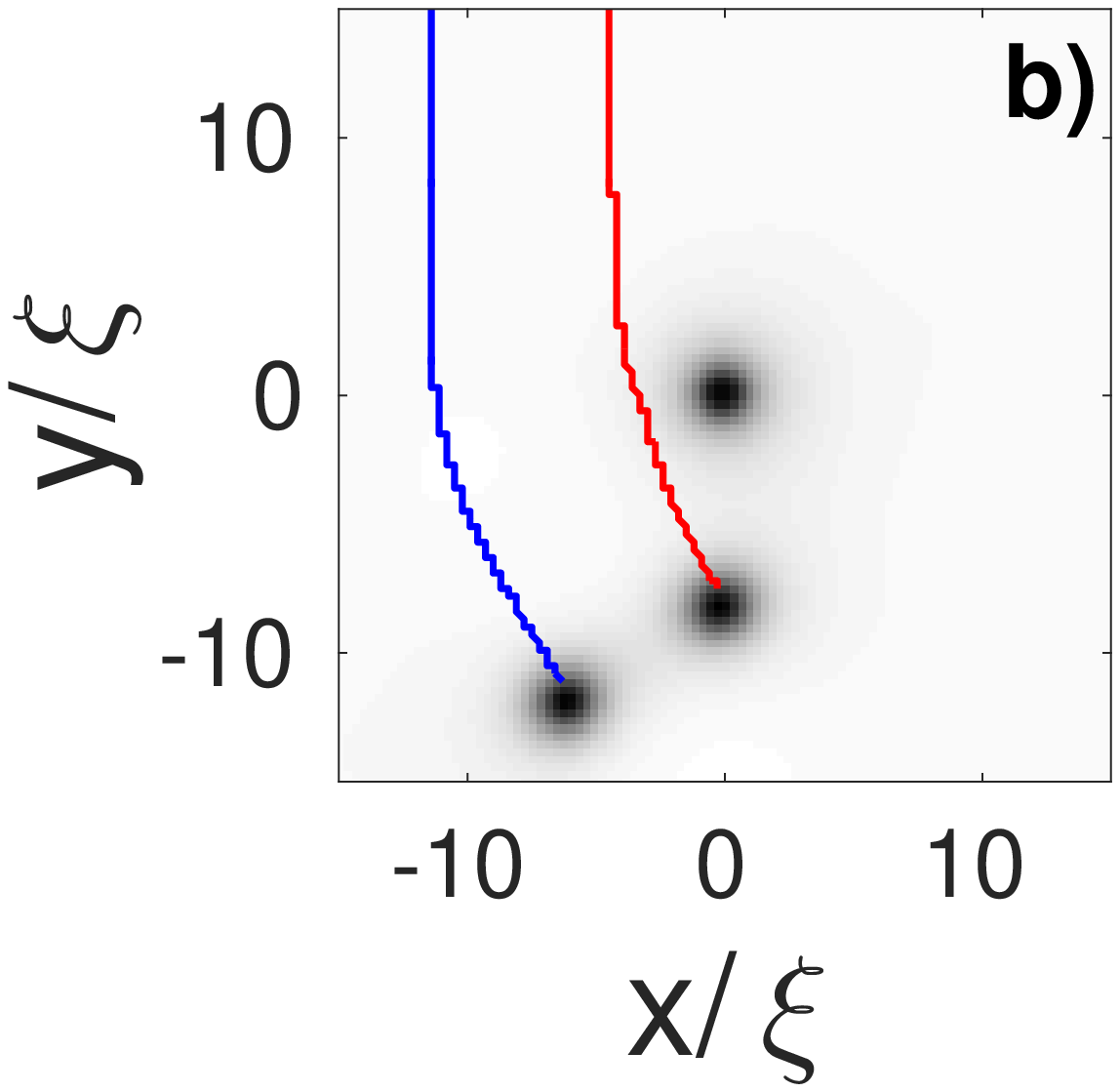}
  \includegraphics[width=0.18\linewidth]{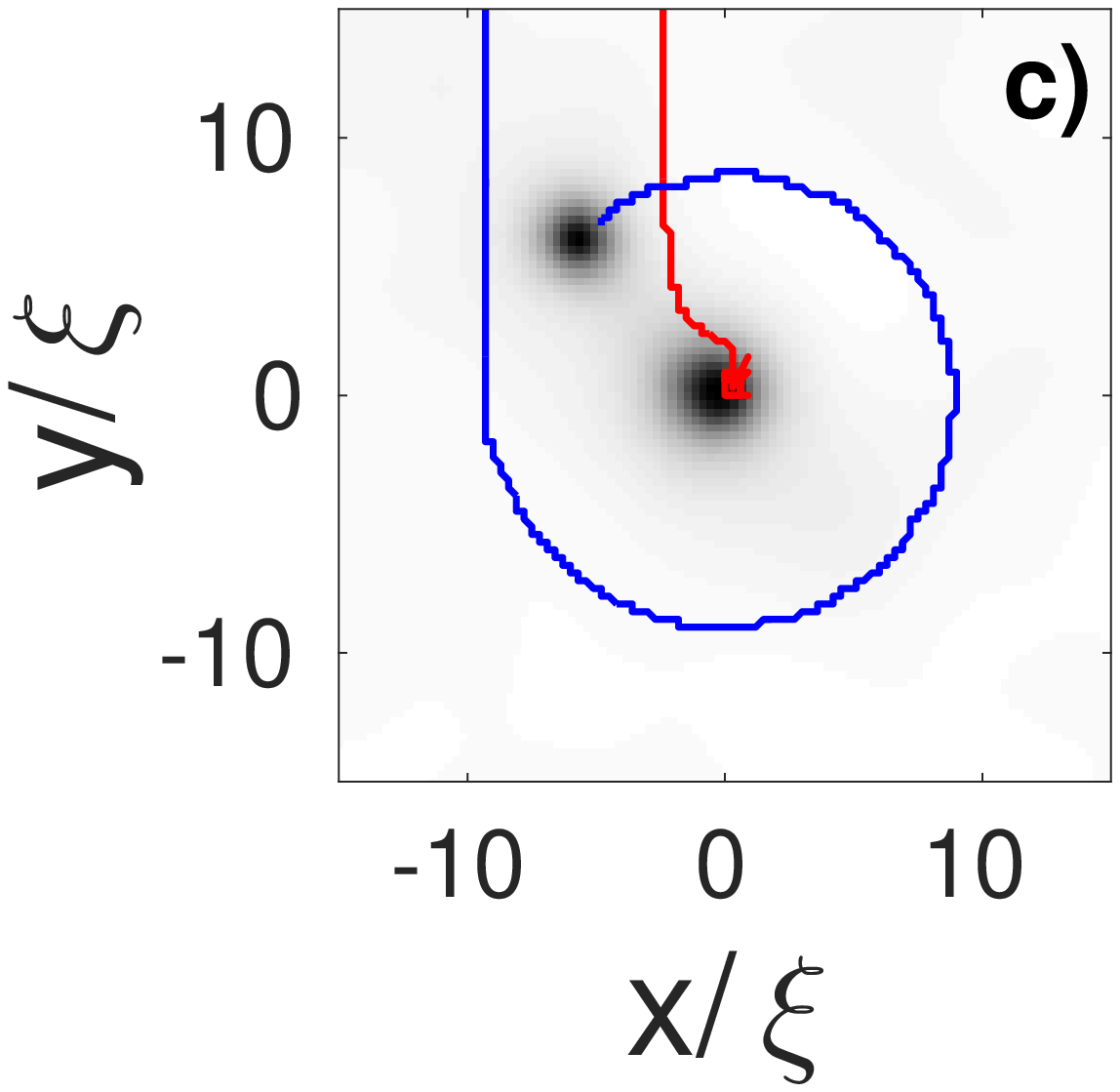}
  \includegraphics[width=0.18\linewidth]{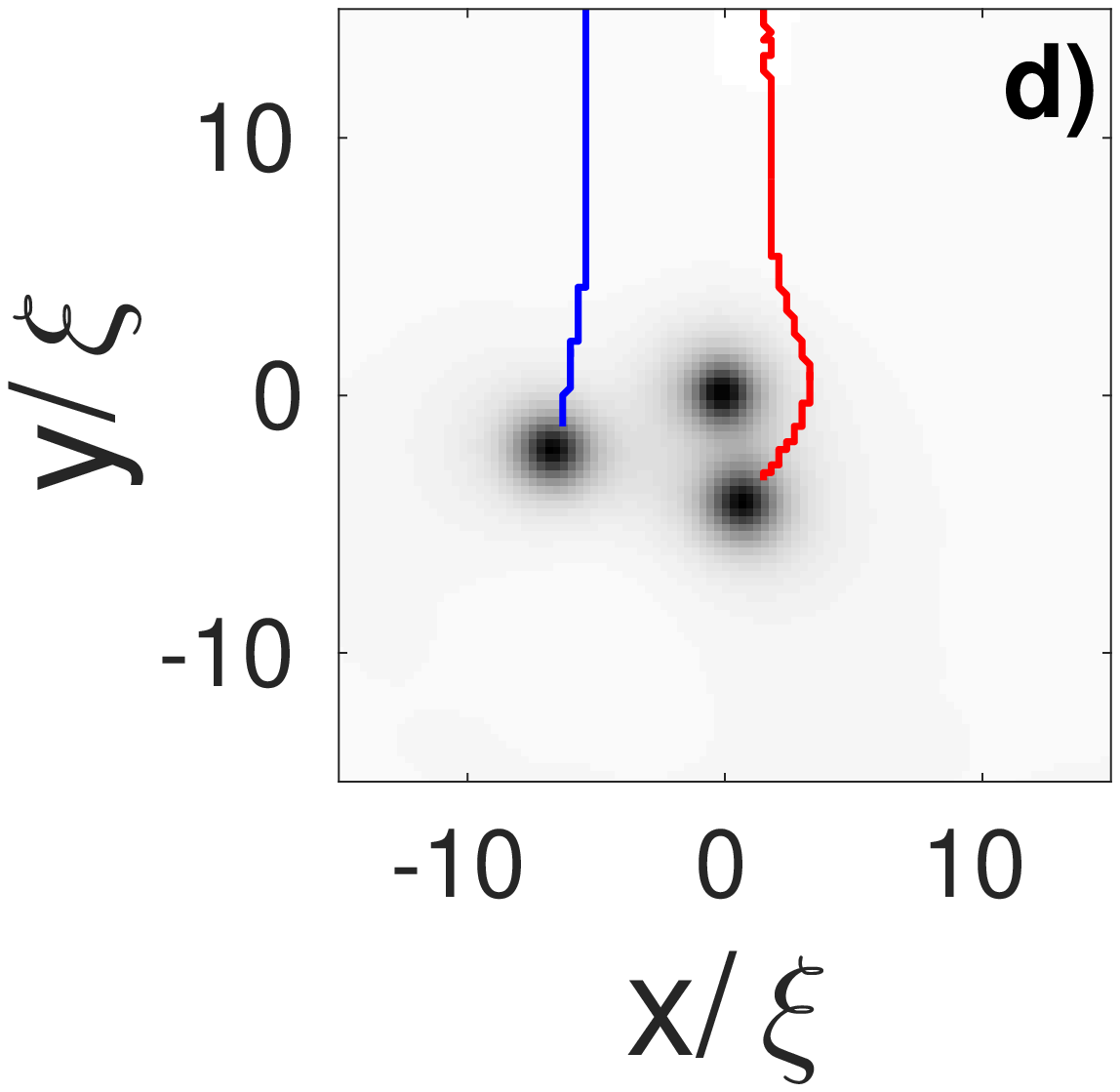}
  \includegraphics[width=0.18\linewidth]{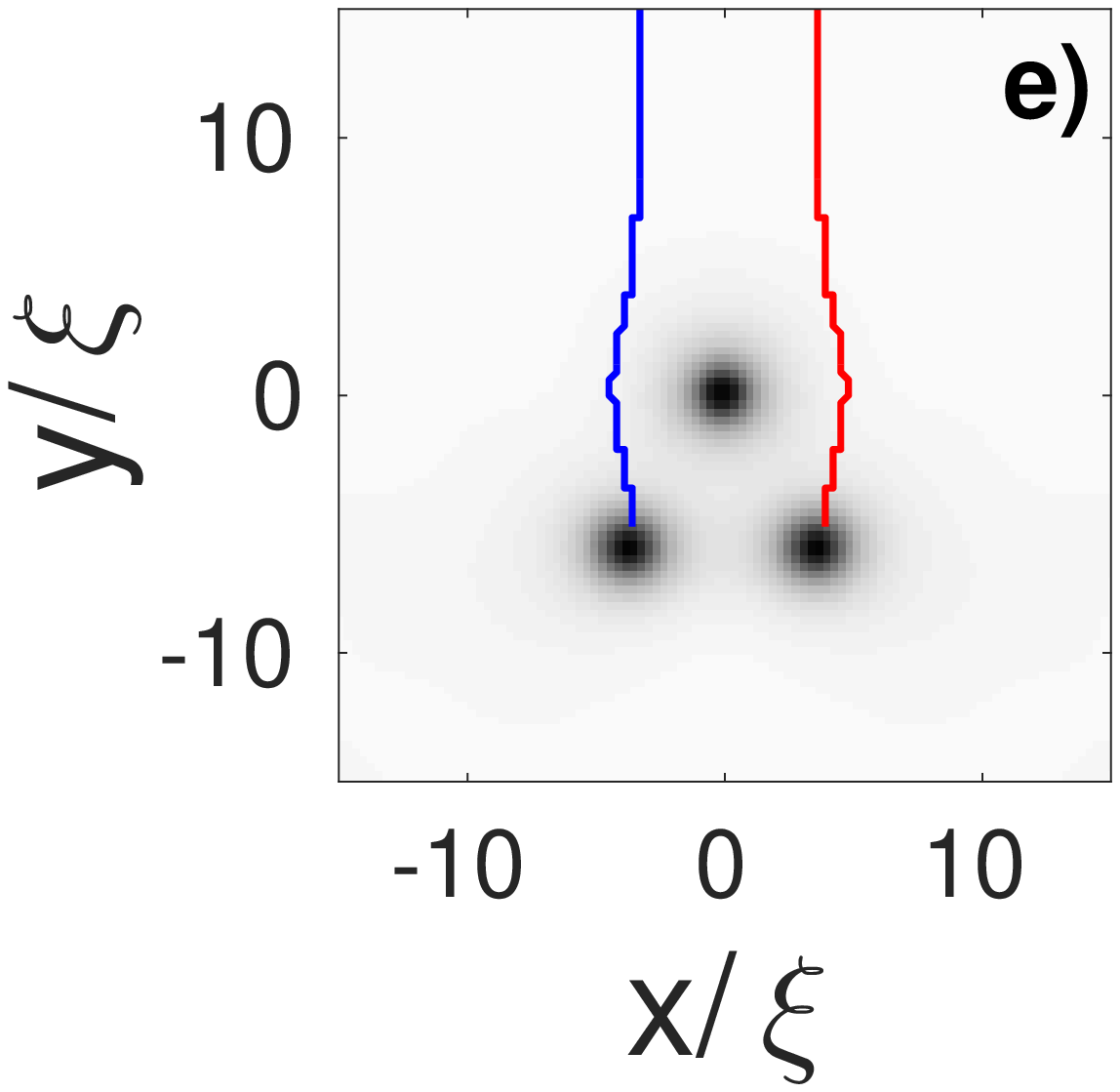}
  \includegraphics[width=0.04\linewidth]{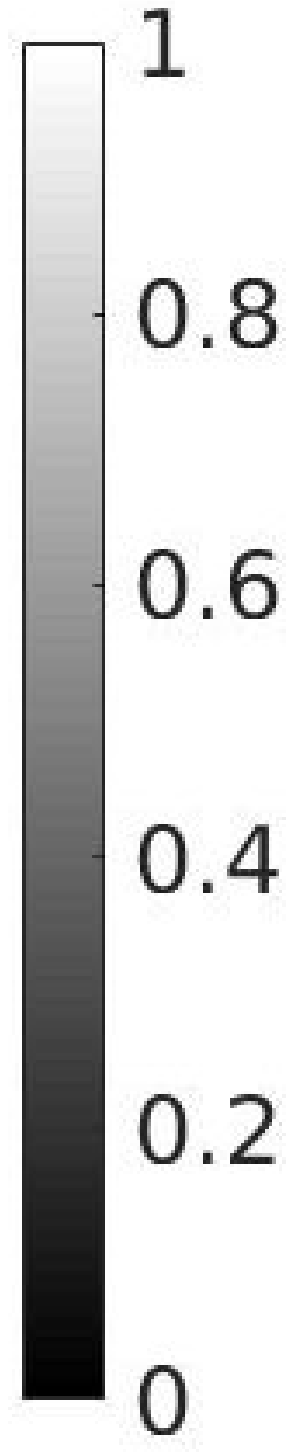}
\caption{(Color online). 
Scenarios of interactions between the vortex-antivortex pair and the
impurity.
In all panels the pair travels from top to bottom toward the impurity; 
the red (right) line and the blue (left) line are the trajectories
of the vortex and the antivortex respectively; 
the black circle at the origin marks the impurity. The top panels
show the computed trajectories; the bottom panels show density
profiles at arbitrarly selected times together with the trajectories.
{\bf (a,b): Fly-by scenario} for $(h,d)=(-9.9\xi, 7.8 \xi)$
and $(-7.8\xi,7.8 \xi)$ respectively. 
The vortex-antivortex pair is scattered by the impurity, deflecting to the 
left by an angle $\theta$.
{\bf (c): Trapping scenario} for $(h,d)=(-5.8 \xi,7,5\xi)$.
The vortex (red trajectory) is trapped by 
the impurity, and the antivortex (blue trajectory) orbits around it.
{\bf (d,e): Go-around scenario} for $(h,d)=(-1.8\xi,7.8 \xi)$ and
$(0.1 \xi, 7.5 \xi)$ respectively.
The vortex and the antivortex overtake the impurity,
going around it in opposite directions.
}
\label{revfig2}
\end{figure}

\begin{figure}[h]
\centering
\includegraphics[width=0.49\linewidth]{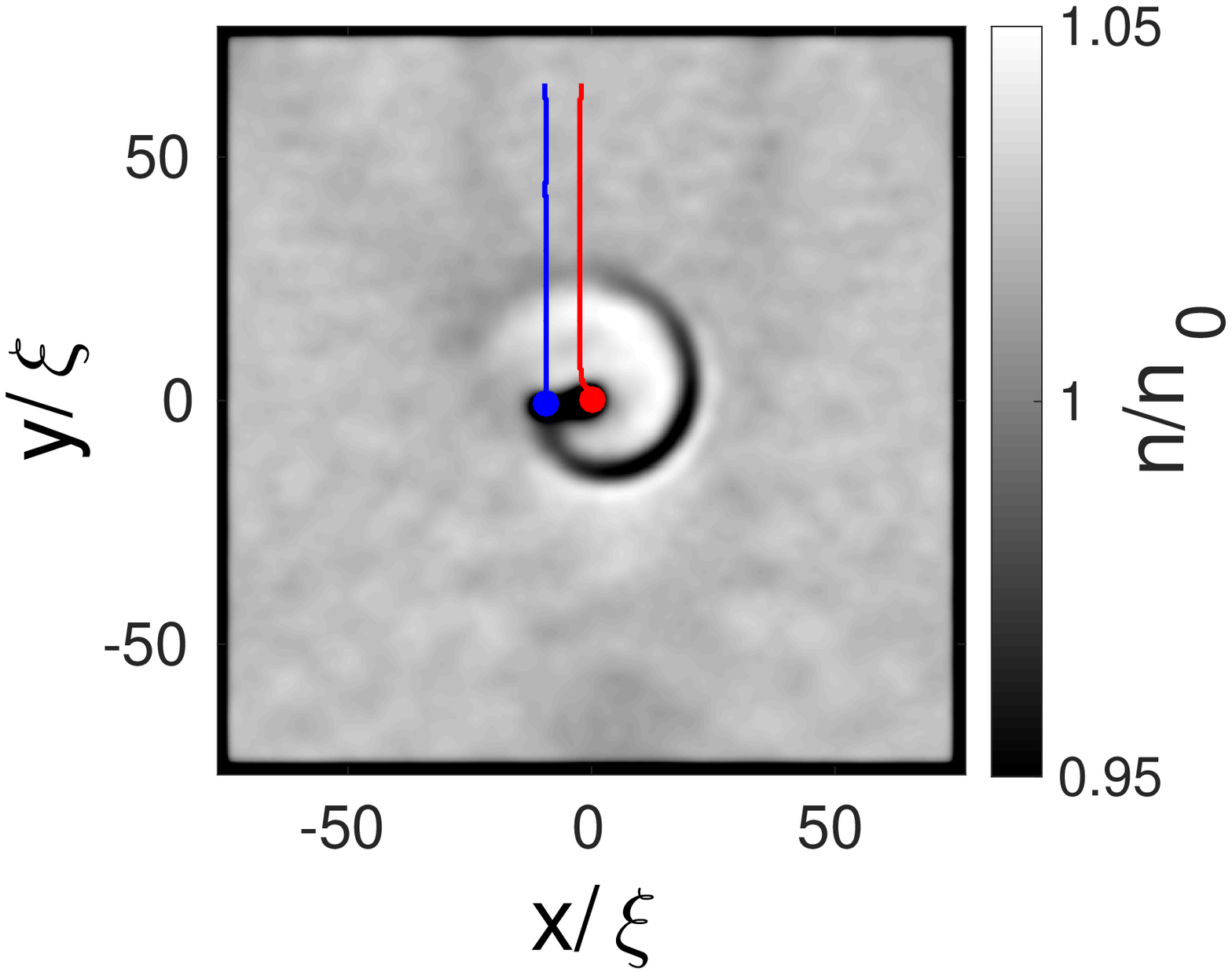}\label{fig:sound2}
\includegraphics[width=0.49\linewidth]{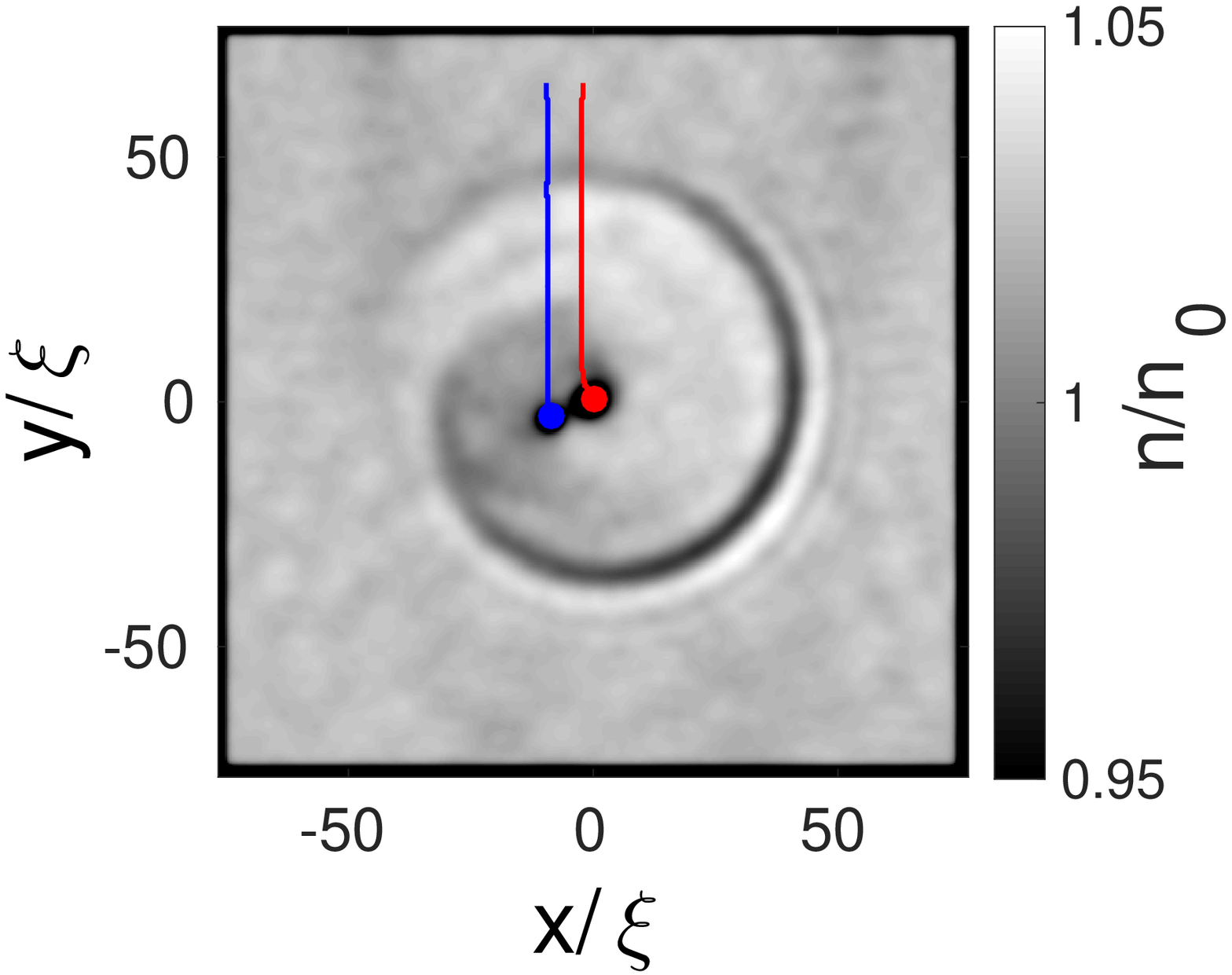}\label{fig:sound3}
 \caption{(Color online).
Trapping scenario. Plots of relative density $n/n_0$  showing the
emission of a sound wave during trapping, corresponding to the
evolution shown in Fig.~\ref{revfig2}(c), at two different times 
$t_1=485 \tau$ (left) and $t_2=505 \tau$ (right). 
The red (right) line
and dot mark the vortex and its trajectory, the blue (left) line and dot
mark the antivortex.
At $t_1$ the vortex and the antivortex are respectively at
$(0,0)$ and $(-9.3\xi,-0.6\xi)$; at $t_2$ they are at
$(0,0.6\xi)$ and $(-8.7\xi,-3.0\xi)$.
}
\label{revfig3}
\end{figure}

In all calculations
we choose initial y-coordinates $y_L(0)=y_R(0)=60 \xi$, 
sufficiently away from the impurity; the initial x-coordinates, $x_L(0)$
and $x_R(0)$, vary from case to case, as we change the impact parameter
$h$ and the vortex separation $d$. 
We have identified three typical scenarios: 

\begin{enumerate}
\item{\bf Fly-by scenario}

If the impact parameter $h$ is large and negative, 
the vortex-antivortex pair is too far at the left of the impurity,
see Fig.~\ref{revfig2}(a), 
to be affected, and the deflection angle
is $\theta \approx 0$.  If the magnitude of $h$ decreases
(still keeping $h<0$),
the vortex-antivortex pair is
scattered to the left (as seen from the initial direction of travel)
with increasing positive deflection angle $\theta$, 
as shown in Fig.~\ref{revfig2}(a,b).

\item{\bf Trapping scenario}

If the magnitude of $h$ is further decreased (still keeping $h<0$), 
the vortex falls into the region of low density of the impurity, see
the red trajectory of Fig.~\ref{revfig2}(c),
becomes trapped  and stops, with
a strong emission of sound waves, see Fig.~\ref{revfig3}; at this
point, the isolated antivortex
processes around the impurity, see the blue trajectory of 
Fig.~\ref{revfig2}(c). 
In this scenario the deflection angle cannot be defined.

\item{\bf Go-around scenario}

A further decrease of the magnitude of $h$ means that the vortex-antivortex
pair is almost aimed at the impurity; the left (anticlockwise) vortex
and the right (clockwise) vortex overtake the impurity on opposite sides,
going around it along opposite directions, before joining again, 
re-forming the pair, and moving on to infinity. 
Fig.~\ref{revfig2}(d) shows that for slightly negative values of
$h$ the vortex pair is scattered to the right ($\theta <0$);  for
$h \approx 0$, see Fig.~\ref{revfig2}(e),
the vortex-antivortex pair proceeds almost straight 
($\theta \approx 0$), vortex and antivortex going around the impurity
in opposite directions.

\end{enumerate}

\begin{figure}
\centering
\includegraphics[width=\linewidth]{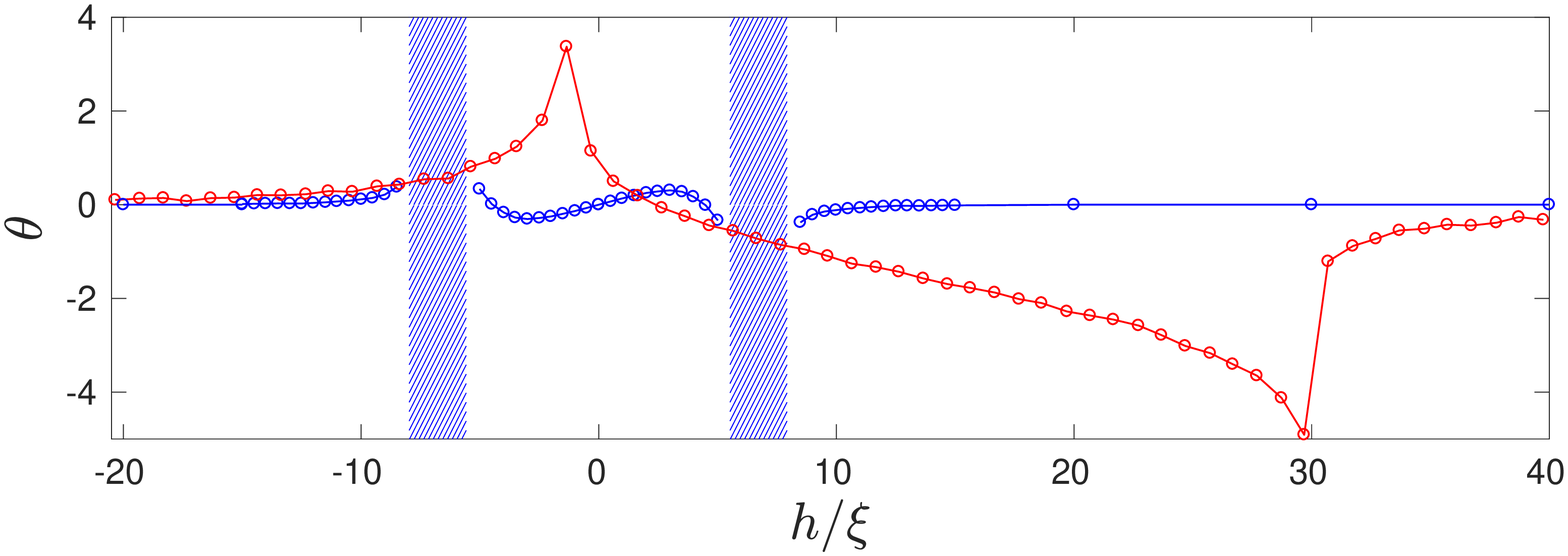}\\
\includegraphics[width=\linewidth]{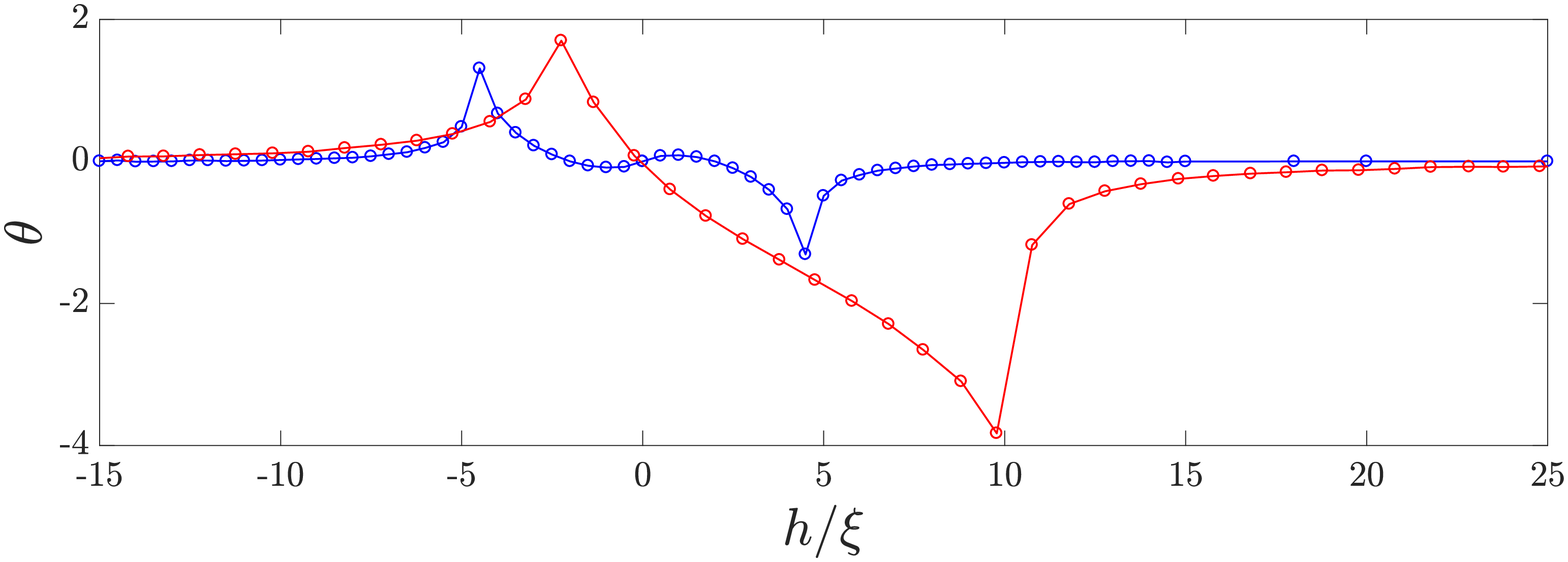}
\caption{(Color online).
Deflection angle $\theta$ as a function of impact parameter $h$ 
for initial vortex-antivortex separation $d=7.8 \xi$ (top)
and $d=3.9 \xi$ (bottom) when the
target is an impurity (blue line and dots) and a vortex (\red{red}
 line and dots).  The shaded blue areas represent the parameter 
regions where $\theta$ cannot be
defined because one vortex becomes trapped in the impurity. 
Comparing top and bottom, notice
the absence of the trapping regime and the extension of the figure to larger
positive values of $h$ (negligible deflection if the target is
an impurity, non-negligible if it is a vortex).
}
\label{revfig4}
\end{figure}

Finally, for larger, positive values of $h$, the trajectories of 
the vortex and the
antivortex are the same (as the impurity does not introduce any preferred
orientation), $\theta$ being replaced by $-\theta$
(in other words the function $\theta(h)$ is antisymmetric in $h$).
We summarize the scenarios which we have revealed
by plotting the deflection angle $\theta$ as 
a function of the impact parameter $h$, see the 
blue line and dots
in Fig.~\ref{revfig4}(top). The shaded
areas represent the regions where the deflection angle $\theta$ cannot be 
defined (one vortex becomes trapped) and the blue line is 
interrupted. 

\begin{figure}
\centering
\includegraphics[width=0.30\linewidth]{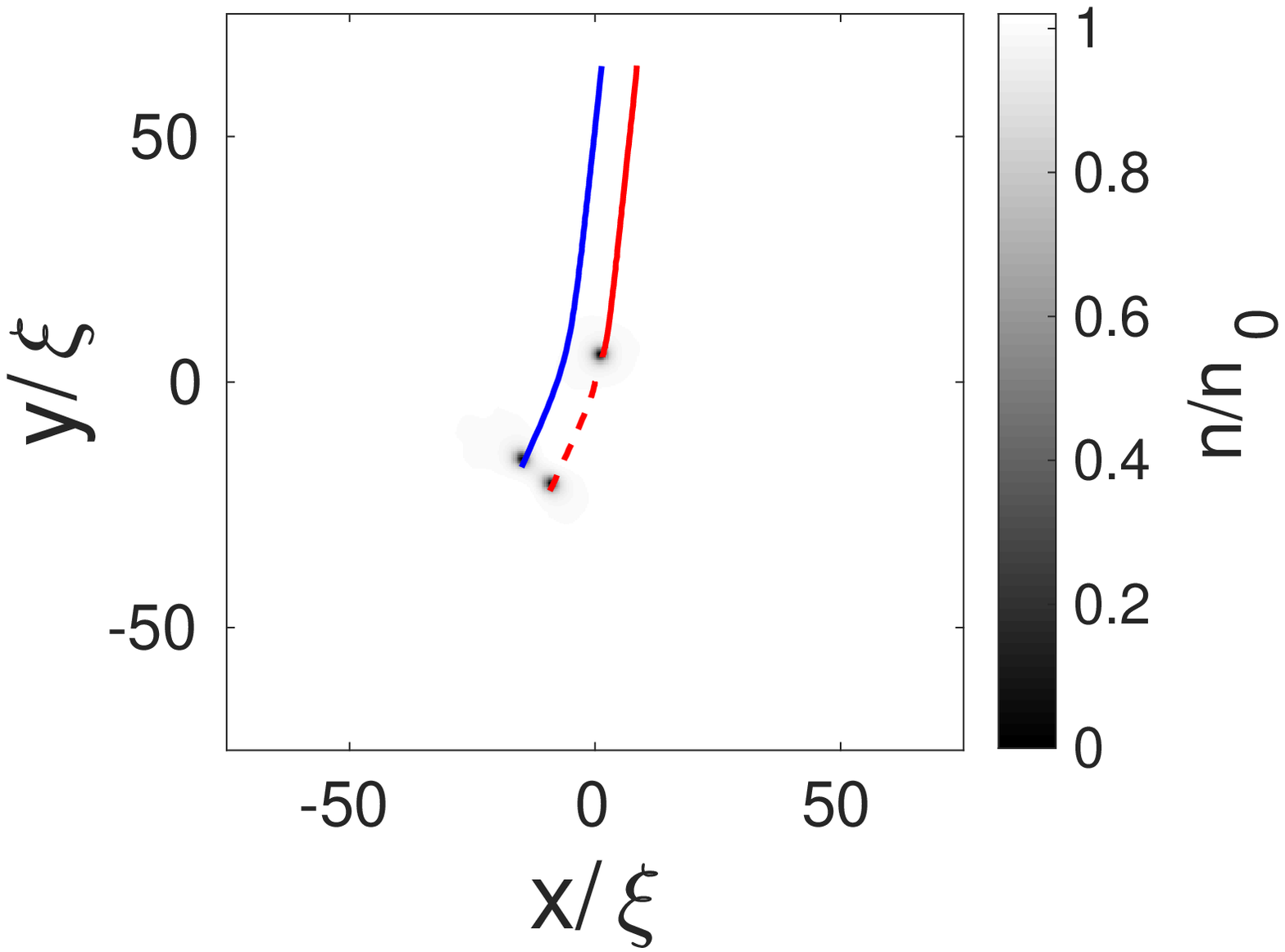}\label{fig:swap}
\includegraphics[width=0.30\linewidth]{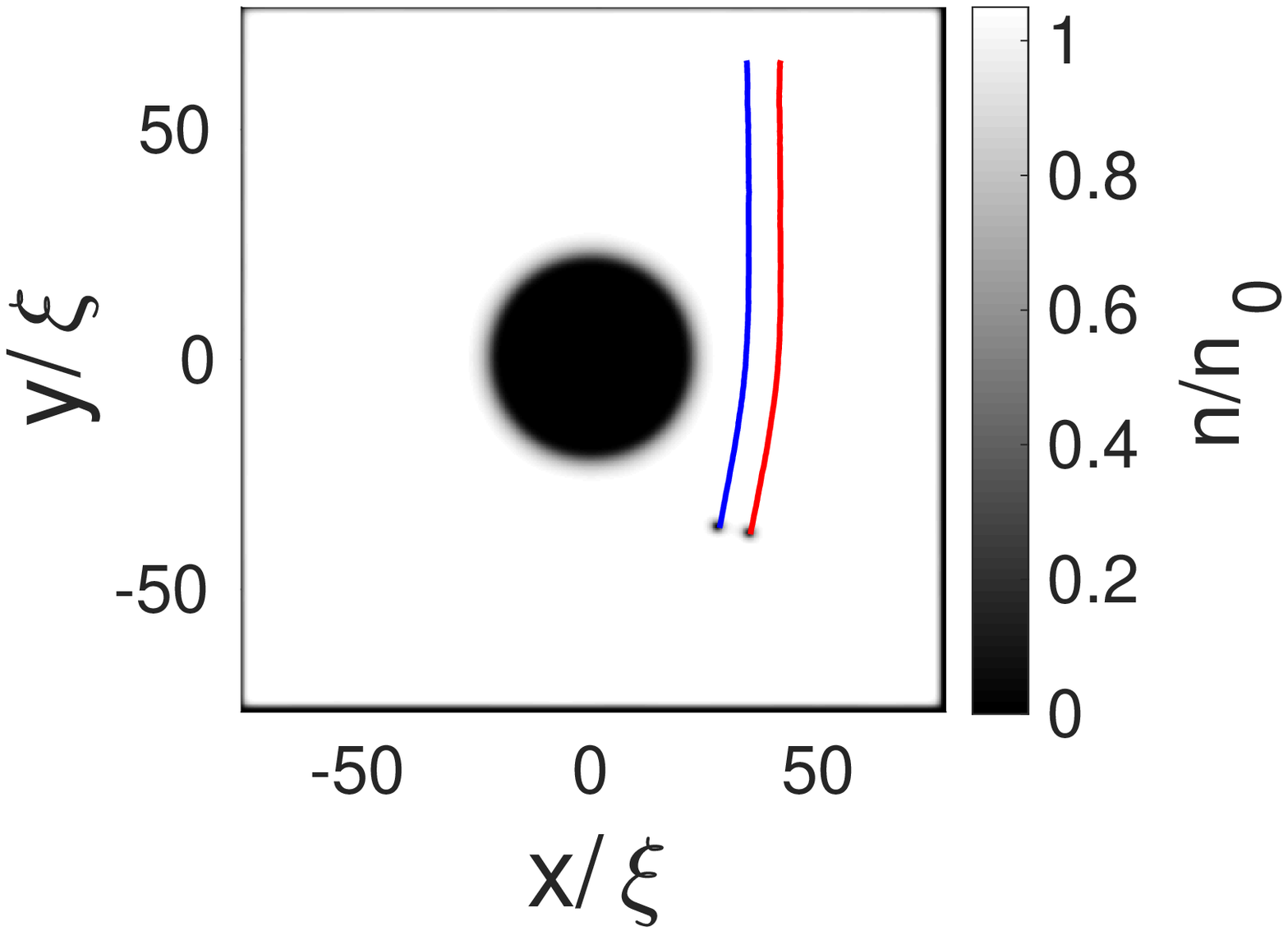}\label{fig:Large_Def}
\includegraphics[width=0.30\linewidth]{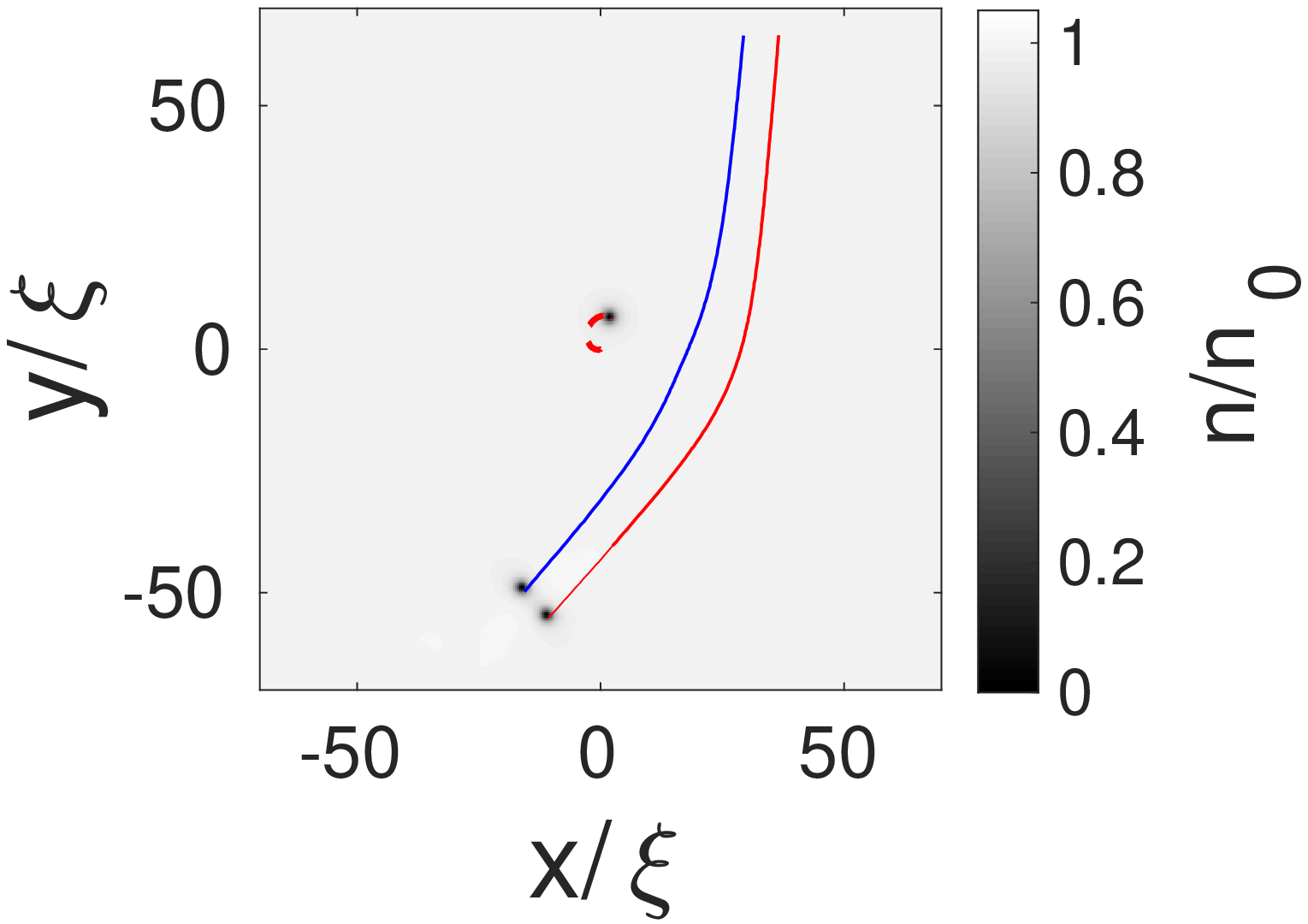}\label{fig:5c}
\caption{(Color online). Left: Example of vortex swapping scenario
(initial separation $d=7.26 \xi$, impact parameter $h=4.66 \xi$). 
Middle: Example of vortex deflection around large impurity.
Right: scattering from a third vortex (initial $d=7.2 \xi$, $h=32.7\xi$);
notice the movement of the third vortex.}
\label{revfig5}
\end{figure}

It is instructive to replace the impurity with a third vortex, 
choosing positive anticlockwise circulation, initially
placed at $x_I=y_I=0$. In this way we can directly compare the deflections of
the vortex pair's trajectory caused by a third vortex to the deflection
caused by an impurity with the same density perturbation, isolating the
effect of the vortex circulation. 
Unlike the impurity, which is fixed, the third vortex
is free to move under the velocity field of the vortex-antivortex pair.
The deflection angle $\theta$ caused by the third vortex is shown by the
red line and dots of Fig.~\ref{revfig4}(top). 
It is apparent that,
for large negative impact parameters, the deflection angle $\theta$
is approximately the same for vortex and impurity, but becomes significantly 
larger for the vortex at small negative $h$; 
moreover, there is no trapping regime
for the vortex. Note also that, for the vortex, the curve $\theta(h)$ is not
antisymmetric \red{about $h=0$} as in the case of the impurity: 
for $h<0$, the closest 
interaction is between vortices of the same sign, which makes the two
close vortices
to rotate around
each other causing a deflection to the left ($\theta>0$) with
respect to the initial direction of motion; for $h>0$,
the closest interaction is between vortices of the opposite sign, which makes
the two close vortices to travel away together, causing a deflection to the 
right ($\theta<0$); with respect to the initial direction of motion. 
This is why the two peaks of the red curve of Fig.~\ref{revfig4}(top),
which represent these strong interactions, are not symmetric about $h=0$.
Between these two peaks there is a regime in which the anti-vortex of the
pair swaps place with the (initially stationary) third vortex, which 
couples with the original vortex of the pair and travels away with it,
forming a new pair.
An example of this swapping regime is presented in Fig.~\ref{revfig5}a.
Finally, notice that the effect of the third vortex extends to 
large positive values of
$h$, unlike the effect of the impurity.

Fig.~\ref{revfig4}(bottom) 
shows what happens if we halve
the initial separation of the vortex from the antivortex
to $d=3.9 \xi$. At this short separation, the trapping
regimes disappears (the vortex, now rather close to the antivortex, moves at
large speed, and the impurity is not strong enough to stop it). The other
features of the interaction remain qualitatively the same as for 
the larger pair separation $d=7.8 \xi$.

\begin{figure}
\centering
\includegraphics[width=\linewidth]{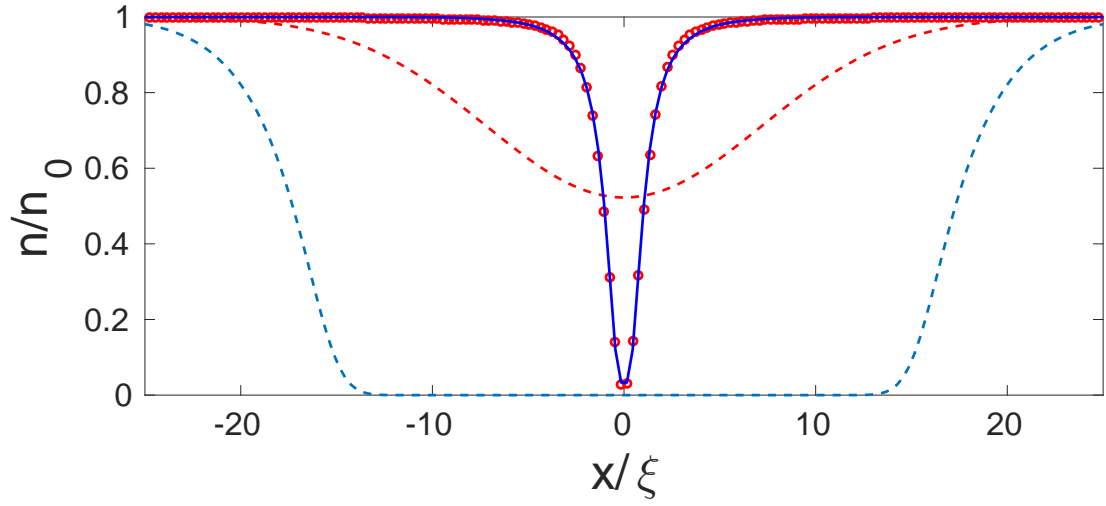}
\caption{(Color online). Relative density profiles, $n/n_0$ vs $x$,
of vortex (solid red line and circles), standard impurity (solid blue line),
large impurity (dashed blue line)
and shallow impurity (dashed red line).
}
\label{revfig6}
\end{figure}

\begin{figure}
\centering
\includegraphics[width=0.49\linewidth]{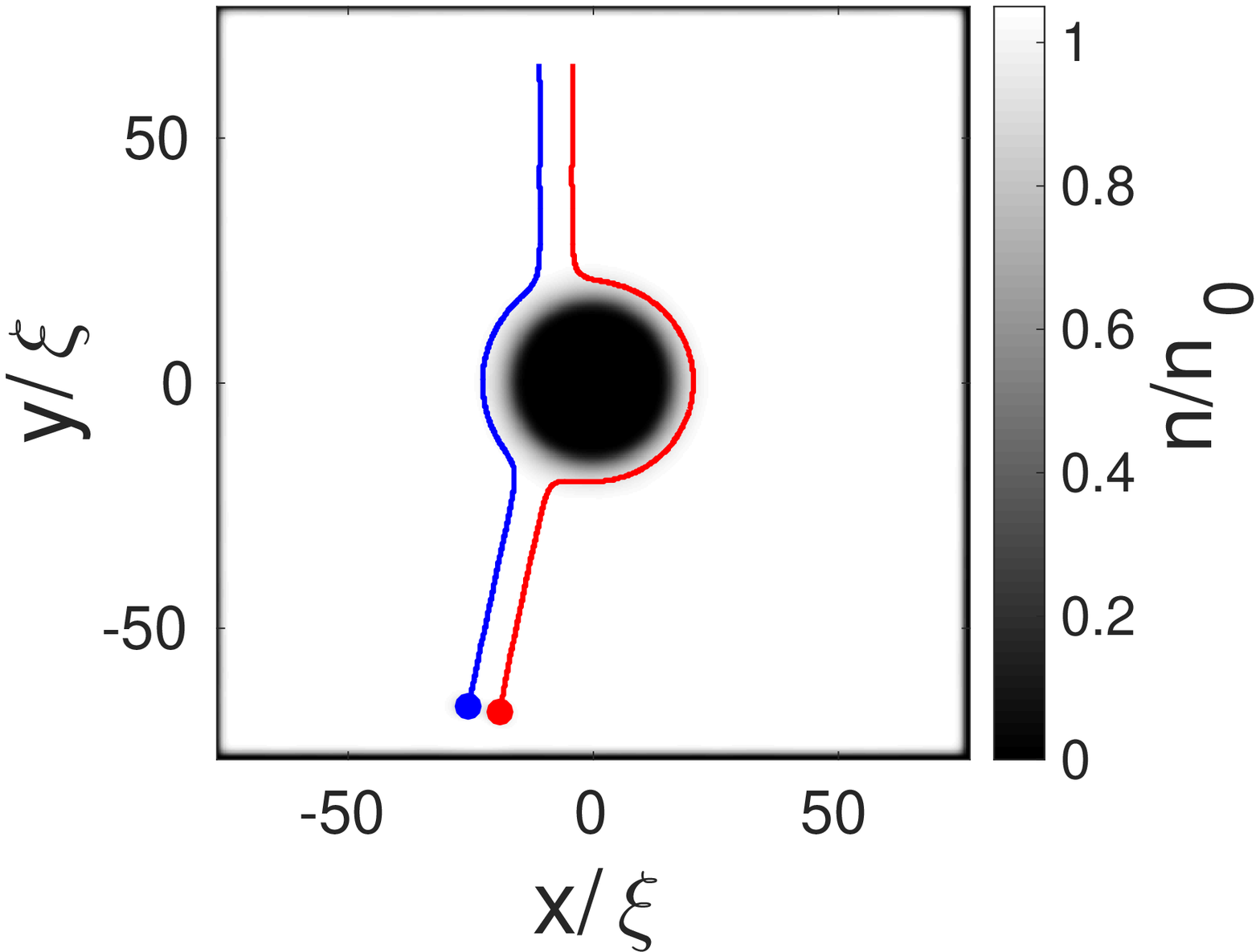}\label{fig:LARGE_V_D}
\includegraphics[width=0.49\linewidth]{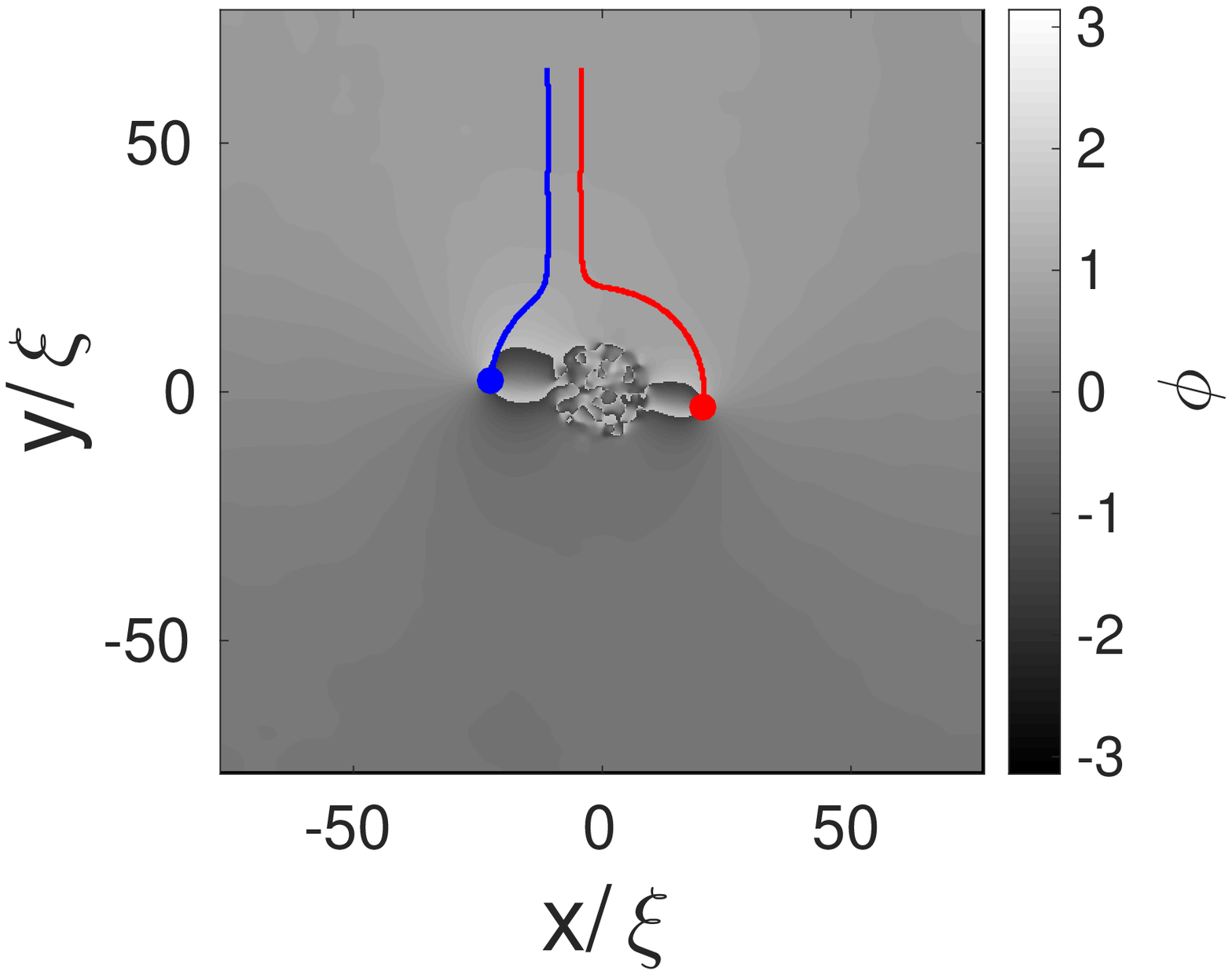}\label{fig:LARGE_V_P}\\
\includegraphics[width=0.49\linewidth]{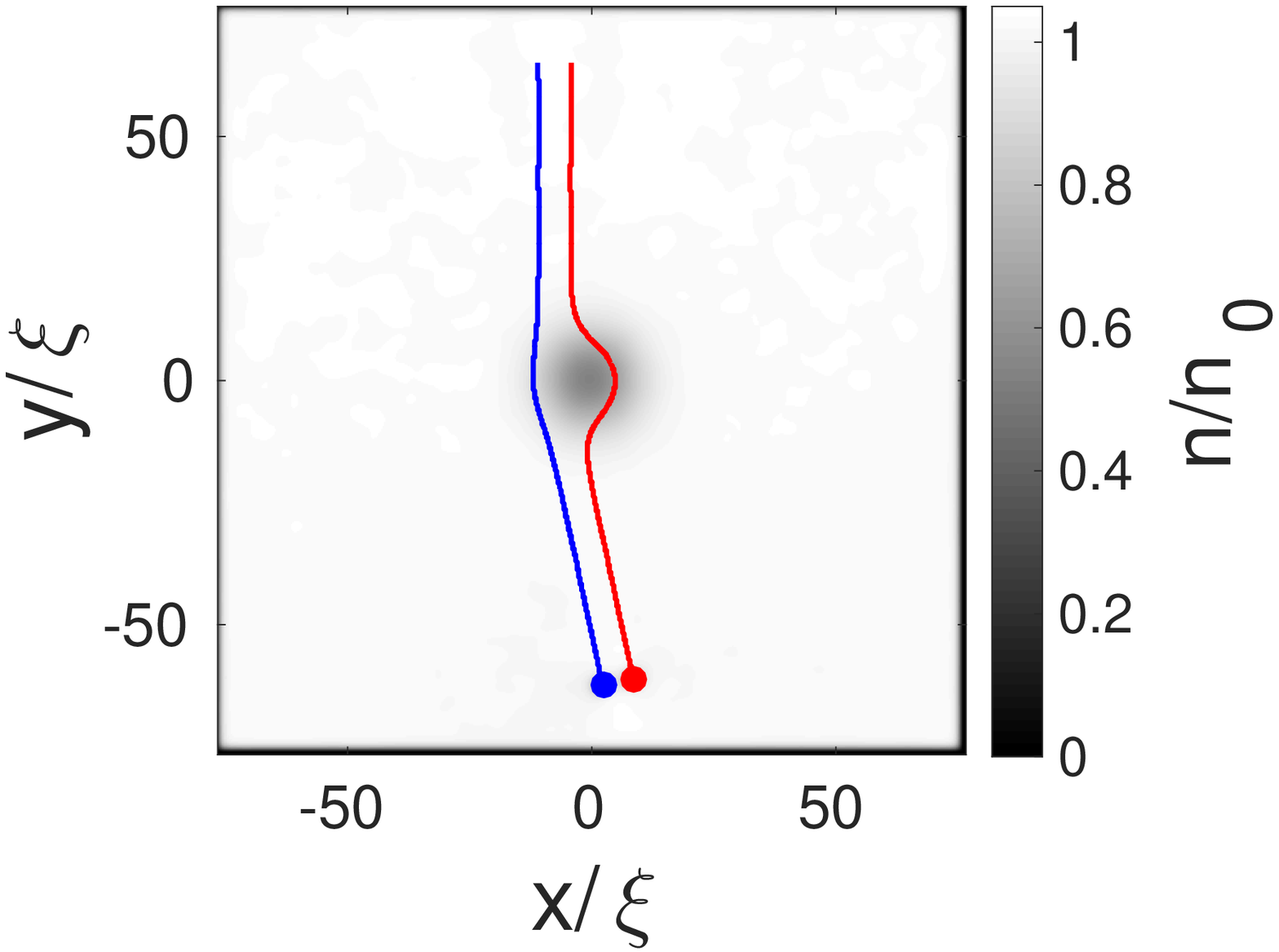}\label{fig:SHALLOW_V_D}
\includegraphics[width=0.49\linewidth]{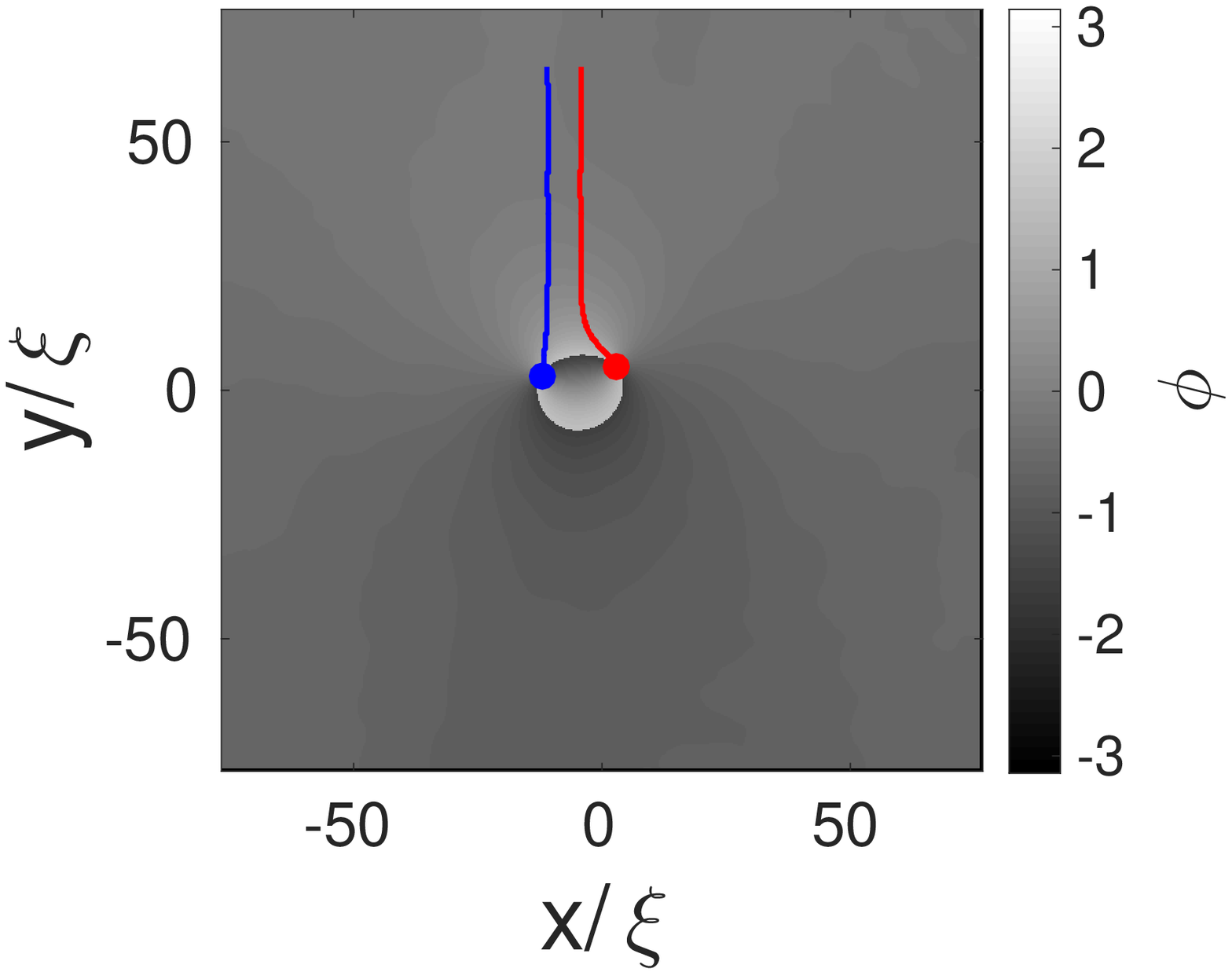}\label{fig:SHALLOW_V_P}
\caption{(Color online).
Go-around scenario for large (top) and shallow (bottom) impurity. 
The trajectories of the vortex
(red line and dot) and the antivortex (blue line and dot) are
superimposed to the density $n(x,y)$ (left) and phase(right).
}
\label{revfig7}
\end{figure}

Increasing the size of the impurity or making it shallower - as for the
density profiles shown by the dashed red and blue lines of
Fig.~\ref{revfig6}  - does not change the scenarios of interaction with
the vortex-antivortex pair in a qualitative way. 
Fig.~\ref{revfig7}(top left) and
Fig.~\ref{revfig7}(bottom left) show the
go-around scenario respectively for a large deep
impurity and for a smaller, shallower
impurity (the density at the centre is only $n \approx 0.5$).
It is interesting to notice that phase defects (ghost vortices) appear
inside the large impurity - compare the top right and bottom right
panels of Figs.~\ref{revfig7}. An example of the 
scattering regime with an impurity similar to the size of the large 
impurity of Fig.~\ref{revfig6} (dashed blue line) is shown in \ref{revfig5}b.

\begin{figure}
\centering
\includegraphics[width=0.99\linewidth]{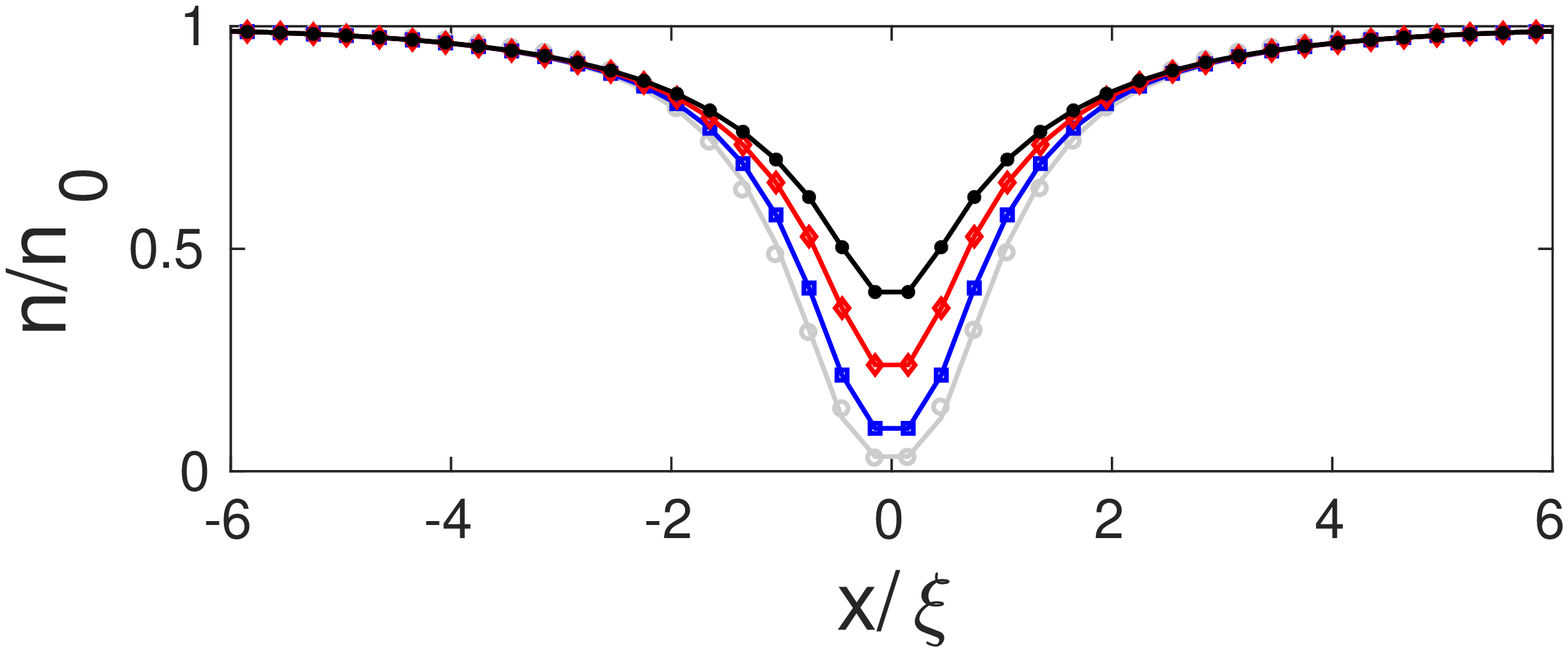}\label{fig:pot_var}\\
\includegraphics[width=0.99\linewidth]{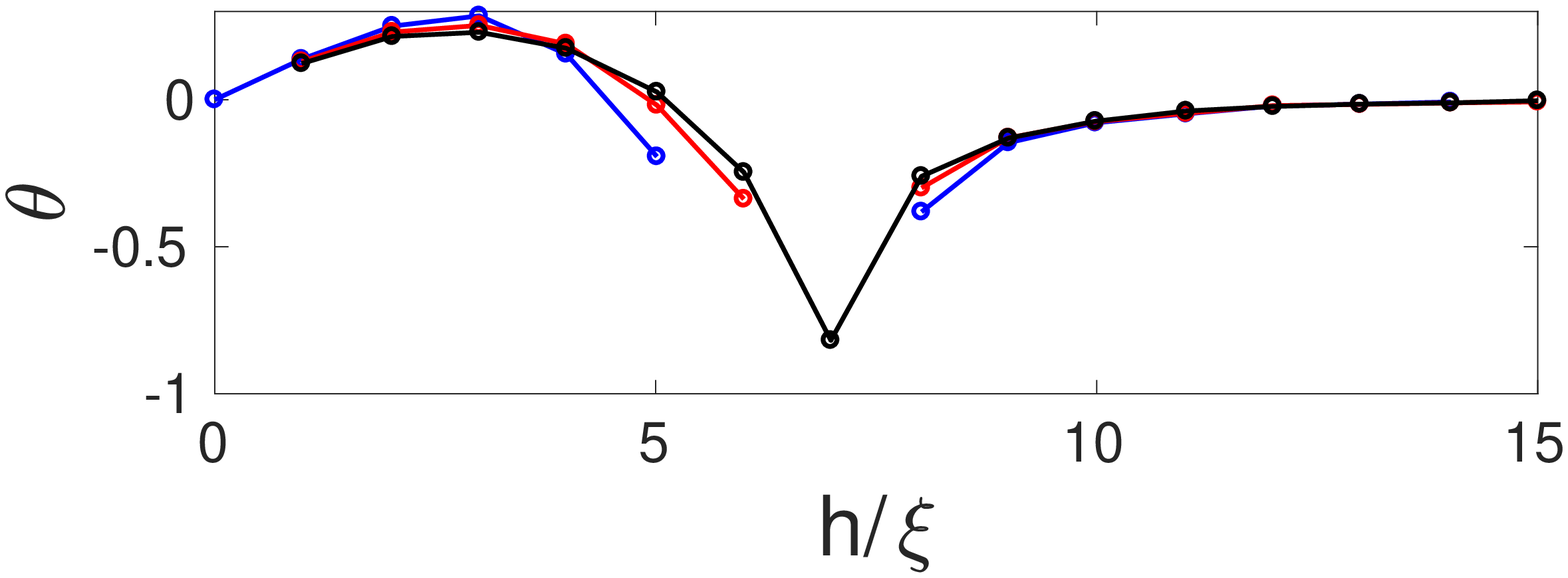}\label{fig:pot_var_de}\\
\caption{(Color online). 
Top: Density profiles, $n/n_0$ vs $x$ for vortex-like potential (gray), 
$n/n_0=0.096$ $A_1=8$ (blue),$n/n_0=0.239$ $A_1=4$ (red) 
and $n/n_0=0.403$ $A_1=2$ (black).
Bottom: Deflection angle $\theta$ against impact parameter $h$. 
Colours correspond to top figure (lines are interrupted in the region
of trapping).}
\label{revfig8}
\end{figure}

Changing the depth of the impurity whilst keeping the width 
close to that of a vortex only modifies the deflection angle slightly.
For deeper impurities we generally see larger scattering angles 
in the region close to the impurity. The general trend is that
the shallower the impurity is, the smaller the region in which trapping 
takes place (see Fig.~\ref{revfig8}), until the impurity is too shallow 
to trap a vortex, as shown by the black line of \ref{revfig8}.

We have already pointed out (Fig.~\ref{revfig3}) that
sound waves created by accelerating vortices \cite{Parker2005}. 
In general, these waves represent small acoustic
losses of kinetic energy which we quantify by recalling the
classical expression for
the energy of a vortex ring of radius $R$ and core radius $a$ in a fluid
of density $\rho$, which is \cite{Barenghi2009}
$E=\rho \kappa^2 R {\cal L}/2$ where
${\cal L}=[\ln{(8R/a)}-2]$. Neglecting variations of the slow
logarithmic term, by measuring the change of distance between the vortex
and the antivortex of the pair, we can estimate the relative energy loss 
$\Delta E/E \approx \Delta R/R$, which is as high as $\Delta E/E \approx 6\%$
for the scattering shown in Fig.~\ref{revfig5}(b).

\section{Discussion and conclusions}

We have compared trajectories of vortex-antivortex pairs launched
either towards a third vortex or toward an impurity in the form of a 
similar density hole but without the circulation. By varying the
impact parameter, we have
identified three general scenarios (fly-by, trapping,
go-around) which can occur. In the first scenario, the effect of the impurity
if qualitative similar to that of the vortex, in the second and third
scenarios it is
significantly different. These scenarios represent the elementary
processes which can be recognized within a 
turbulent system. They are therefore relevant to experiments
in which vortices are manipulated by laser beams and to studies of
2D quantum turbulence, as large density perturbations are often
generated by vortex annihilations or by the moving laser beam used to
nucleate vortices in the first place. 

Our results are consistent with work on the scattering of 2D
quasi-solitons from potential barriers \cite{MironovSmirnov2012}, for
example we observe that a vortex pair is deflected towards a density dip 
rather than away from it as predicted (see their Fig.~1, bottom).
Our work is motivated by the aim of getting insight into what is typically 
seen in experiments and numerical simulations of 2D vortex turbulence, 
and differs from Ref.~\cite{MironovSmirnov2012} in three respects. 
Firstly it is concerned with well-separated vortex and antivortex rather
than solitons (when the speed of the pair exceeds the 
critical value $v/c=0.61$ where $c$ is the sound speed, the circulation
is lost and the pair becomes a solitonic object). Secondly it refers
to much smaller impurities (of the order of the core size, not ten times
larger).  In particular, unlike Ref.~\cite{MironovSmirnov2012}, in our work
the vortex and the antivortex which make up a pair can separate. Thirdly,
by solving directly the GPE, we allow acoustic losses unlike the model
equations of Ref.~\cite{MironovSmirnov2012}.

Future work should look at the
effects induced by an inhomogeneous density background in
harmonically trapped condensates. Other aspects which are worth
investigating are thermal and quantum fluctuations, which are
not included in our mean field GPE model.
Qualitatively, one would expect thermal fluctuations to move
the vortex and the antivortex of a pair closer to each other,
eventually leading to their annihilation.
This effect would lead to the introduction 
of a new length/time scale associated with the vortex-antivortex pair's
intrinsic decaying dynamics. Qualitatively, however, we would still 
expect the same regimes to emerge. Quantum 
fluctuations would lead to an intrinsic jitter motion of each vortex
about its mean position, with the target impurity also suffering some 
fluctuations from the fluctuating density in that region.
On the average we would still expect the fly-be and go-around 
scenarios to persist, but our discussion here may represent a rather 
idealised case. In fact it would be interesting to investigate 
this scenario experimentally.

\begin{acknowledgements}
We acknowledge the support of EPSRC grants EP/K03250X/1 and EP/I019413/1.
\end{acknowledgements}

\newpage

\vskip 4cm
\vfill
\eject

\newpage

\vfill
\eject

\newpage

\vfill
\eject

\newpage

\vfill
\eject

\newpage

\vfill
\eject

\newpage

\vfill
\eject

\newpage

\vfill
\eject

\newpage

\end{document}